\documentclass[conference, a4paper, onecolumn]{IEEEtran}
\IEEEoverridecommandlockouts
\usepackage[left=1.5cm,right=1.5cm,top=2cm,bottom=4.4cm]{geometry}
\usepackage{mleftright}       % fix to wrong spacing of \left-,
\mleftright                   % \middle- \right-commands 

\usepackage{graphicx}         % provides \includegraphics{...} to
\usepackage{booktabs} 

\usepackage[utf8]{inputenc}
\usepackage{amsmath,amssymb,amsfonts}
\usepackage{mathtools}
\usepackage{amsthm}
\usepackage{mathrsfs}
\usepackage{algorithmic}
\usepackage{ragged2e}
\usepackage{textcomp}
\usepackage{tikz}
\usepackage{caption}
\usepackage{cuted}
\usepackage{pgfgantt}
\usepackage{pdflscape}
\usepackage{pst-plot}
\usepackage{comment} 
\usepackage{cases}
\usepackage{nccfoots}
\usepackage{lineno}
\usepackage{graphicx}
\usepackage{float}
\usepackage[caption=false,font=footnotesize]{subfig}
\usetikzlibrary{spy}
\usetikzlibrary{positioning,calc}
\usetikzlibrary{decorations.pathmorphing,calc,shapes,shapes.geometric,patterns}
\usetikzlibrary{shapes.multipart}
\usepackage{xfrac}
\usepackage{colortbl}
\usepackage{cancel} 
\usepackage{bbm}
\usetikzlibrary{arrows,positioning,calc,intersections}
\usetikzlibrary{datavisualization.formats.functions}
\def\BibTeX{{\rm B\kern-.05em{\sc i\kern-.025em b}\kern-.08em
    T\kern-.1667em\lower.7ex\hbox{E}\kern-.125emX}}
    \newcommand{\norm}[1]{\left\lVert#1\right\rVert}
\usepackage{romannum}
\usepackage{pgfplots}
\usepgfplotslibrary{fillbetween}
\usetikzlibrary{arrows, decorations.markings}
\usetikzlibrary{arrows.meta}

\newtheorem{theorem}{Theorem}
\newtheorem*{theorem*}{Theorem}
\newtheorem{lemma}[theorem]{Lemma}

\newtheorem{definition}[theorem]{Definition}

\renewcommand{\baselinestretch}{1}
\DeclarePairedDelimiterX{\infdivx}[2]{(}{)}{%
  #1\;\delimsize\|\;#2%
}

\newcommand\blfootnote[1]{%
  \begingroup
  \renewcommand\thefootnote{}\footnote{#1}%
  \addtocounter{footnote}{-1}%
  \endgroup
}

\pgfplotsset{compat=1.18}
\begin{document}
\title{Identification over Poisson ISI Channels: Feedback and Molecular Applications} 

% %%% Single author, or several authors with same affiliation:
% \author{%
%   \IEEEauthorblockN{Stefan M.~Moser}
%   \IEEEauthorblockA{ETH Zürich\\
%                     ISI (D-ITET)\\
%                     CH-8092 Zürich, Switzerland\\
%                     Email: moser@isi.ee.ethz.ch}
% }

%%% Several authors with up to three affiliations:

\author{
\IEEEauthorblockN{
Yaning Zhao\IEEEauthorrefmark{1}\textsuperscript{,1}, 
Pau Colomer \IEEEauthorrefmark{2}\textsuperscript{,1,2}, 
Holger Boche\IEEEauthorrefmark{2}\textsuperscript{,1,3,4,5}, 
and Christian Deppe\IEEEauthorrefmark{1}\textsuperscript{,1}}
\IEEEauthorblockA{\IEEEauthorrefmark{1}Technical University of Braunschweig,
\IEEEauthorrefmark{2}Technical University of Munich\\
Email: yaning.zhao@tu-bs.de, pau.colomer@tum.de, boche@tum.de, christian.deppe@tu-bs.de}
}

\maketitle

%%%%%% 
%% Abstract: 
%% If your paper is eligible for the student paper award, please add
%% the comment "THIS PAPER IS ELIGIBLE FOR THE STUDENT PAPER
%% AWARD." as a first line in the abstract. 
%% For the final version of the accepted paper, please do not forget
%% to remove this comment!
%%

%====================== abstract =========================
\begin{abstract}
Molecular communication (MC) enables information transfer via molecules, making it ideal for biomedical applications where traditional methods fall short. In many such scenarios, identifying specific events is more critical than decoding full messages, motivating the use of deterministic identification (DI). This paper investigates DI over discrete-time Poisson channels (DTPCs) with inter-symbol interference (ISI), a realistic setting due to channel memory effects. We improve the known upper bound on DI capacity under power constraints from $\frac{3}{2} + \kappa$ to $\frac{1 + \kappa}{2}$. Additionally, we present the first results on deterministic identification with feedback (DIF) in this context, providing a constructive lower bound. These findings enhance the theoretical understanding of MC and support more efficient, feedback-driven biomedical systems.
\end{abstract}
%\textit{A detailed version with all proofs, explanations, and more discussions can be found in {\color{red} [perhapsanarxivversion]}}.

%============== additional affiliations ==================
\blfootnote{
{\textsuperscript{1}BMBF Research Hub 6G-life, Germany.}

{\textsuperscript{2}Institute for Advanced Study (IAS-TUM), Lichtenbergstr. 2, Garching, Germany.}

{\textsuperscript{3}Munich Center for Quantum Science and Technology (MCQST), Schellingstra{\ss}e 4, München, Germany.}

{\textsuperscript{4}Munich Quantum Valley (MQV), Leopoldstra{\ss}e 244, M\"unchen, Germany.}

{\textsuperscript{5}Cyber Security in the Age of Large-Scale Adversaries–Exzellenzcluster, Ruhr-Universit\"at Bochum, Germany.}
}

%===================== introduction ======================

\section{Introduction}
\label{sec:intro}
%====== MC =======
Molecular communication (MC) has emerged as a groundbreaking paradigm in communication technology, leveraging molecules as carriers of information instead of conventional electromagnetic waves. This biologically inspired approach is particularly promising for applications in biotechnology and medicine, where traditional communication methods face significant limitations \cite{nakano2013molecular, nakano2012molecular}. Indeed, MC has been widely explored for its potential in targeted drug delivery \cite{muller2004challenges}, disease detection \cite{ghavami2017abnormality,mosayebi2017cooperative,ghoroghchian2019abnormality}, and real-time health monitoring \cite{nakano2012molecular}. 

In an MC system, information is transmitted through the controlled release of molecules by a sender, such as a biological cell or engineered nanomachine \cite{nakano2013molecular}. These molecules propagate through the medium via diffusion or advection, potentially undergoing enzymatic degradation before reaching the receiver \cite{farsad2016comprehensive}. The receiver, which could be another cell or a nanoscale biosensor, detects the transmitted molecules and interprets the signal based on their quantity. 
%The stochastic nature of molecular propagation introduces unique challenges in modeling and optimizing MC systems. 
One widely accepted framework for analyzing MC channels is the discrete-time Poisson channel (DTPC), which captures the statistical properties of molecular arrivals when a large number of molecules are released \cite{ghoroghchian2019abnormality,jamali2019channel,unterweger2018experimental}. However, constraints on molecular production and release rates impose fundamental limits on the system's performance, necessitating careful consideration of average and peak power constraints \cite{jamali2019channel,gohari2016information}.

%===== ID =========
While traditional communication theory emphasizes the decoding of transmitted messages, many MC applications are inherently event-driven \cite{cabrera20216g}, such as targeted drug delivery \cite{manish2011targeted}, environment monitoring \cite{qiu2014molecular}, and advanced cancer therapies \cite{wilhelm2016analysis}. In such scenarios, the receiver’s primary objective is not to reconstruct the transmitted information in full but rather to determine whether a specific event has occurred. This shift in focus aligns with the identification (ID) scheme, initially introduced by Ahlswede and Dueck \cite{ahlswede1989identification}. Previous research has demonstrated that for discrete memoryless channels (DMCs) \cite{ahlswede1989identification}, the number of identifiable messages grows doubly exponentially when randomized encoding is permitted, i.e., $\sim2^{2^{nR}}$, where $R$ is the coding rate. Similarly, deterministic ID (DI) over memoryless channels with finite output spaces and arbitrary input alphabets has been studied in \cite{colomer2025deterministic,colomer2025ratereliability}, demonstrating that the number of identifiable messages exhibits super-exponential growth $\sim 2^{n\log(n)R}$. Specifically for channels relevant for MC, DI has been studied for binomial channels in \cite{salariseddigh2023deterministic}, for DTPCs in \cite{dIDMC}, and for fading channels in \cite{salariseddigh2021deterministic}. Additionally, randomized ID over DTPCs has been investigated in \cite{10925939}. 

%\pau{Specially in the MC setting, isn't it more relevant the message length ($\log M$) than the actually number of transmittable or identifiable messages ($M$). The vocabulary would then change from exponential $\rightarrow$ linear, from super-exponential $\rightarrow$ superlinear, and from double exponential $\rightarrow$ exponential. Matching in this way the vocabulary in \cite{colomer2025deterministic}, and the subsequent reliability, steins, and quantum hypothesis testing works.} \yz{This part I am not very sure. @Christian: What do you think? }
%\cd{With Holger and Uzi I used superexponential. Later when Andreas and Pau joined we changed because of the argument (see above). For me both is fine.}
%\pau{I guess that the decision is rather which magnitude do we fine more interesting to describe in this setting, the message length $\log M$ or the number of different messages $M$? Specifically for DNA I would argue that the message length might be more relevant, but at the end of the day it is a matter of taste}
% =========== feedback ===========
Progress in ID schemes can leverage resources such as quantum entanglement \cite{boche2019secure,pereg2022identification}, common randomness (CR) \cite{ezzine2024common,9517972,labidi2022common}, and feedback \cite{IDF}. Particularly, feedback could have significant applications in MC systems \cite{6179346,moore2007interfacing}, especially in medical treatment \cite{felicetti2016applications,moritani2006molecular} and precisely controlling cellular processes using injected nanomachines \cite{6708564}. Imagine a patient in a CT or MRI scanner, where a nanomachine is injected to interact with a target cell and trigger actions like protein release or repair mechanisms \cite{kong2023advances}. The nanomachine communicates with the cell through signaling molecules, and advanced imaging technologies could monitor how many molecules are absorbed in real-time. This feedback allows the nanomachine to adjust its behavior dynamically, optimizing timing, dosage, and type of molecules released \cite{jones2019towards,kim2018artificial}. This process aligns with the DTPC model with feedback, where the nanomachine adapts based on the cell’s response, enabling more targeted and efficient treatments.

%============ ISI =================

In MC systems, inter-symbol interference (ISI) and channel memory are inherent to the transmission process \cite{mahfuz2011characterization}. Molecules persist in the medium, influencing the reception of subsequent symbols, creating a natural form of interference. Incorporating these effects is essential for realistically modeling MC, especially in complex environments. The work in \cite{pierobon2012capacity} addresses the transmission capacity of MC with channel memory and molecular noise. In \cite{dIDMCISI, salariseddigh2024deterministic}, the investigation of DI capacity and K-identification (K-ID) capacity for DTPCs with ISI has been explored.

%========== contributions ==========

In this work, inspired by the converse proof in \cite{colomer2025deterministic} for DI over channels with finite output, we significantly improve the upper bound of the DI capacity for the DTPC channel with ISI under power constraints. Specifically, we reduce it from $\frac{3}{2} + \kappa$ (the result in \cite{dIDMCISI}) to $\frac{1 + \kappa}{2}$ under both average and peak power constraints. Furthermore, to the best of our knowledge, we address for the first time the deterministic identification with feedback (DIF) problem over DTPC channels with ISI under power constraints, providing a specific coding scheme to characterize a lower bound on the DIF capacity. As it turns out, in the presence of feedback, the DIF code sizes immediately transition to a double exponential scale (like in the randomized scenario). This shows the value of feedback in ID compared to Shannon's classical transmission scheme, where feedback provides no relevant advantage.

\textit{Outline}: The remainder of the paper is structured as follows. In Section \ref{sec:main}, we introduce our system model and present the main results. In Section \ref{sec:proof}, we prove the improved upper bound on the DI capacity and the novel lower bound on the IDF capacity. Finally, in section \ref{sec:conclusion}, we provide some conclusions and new possible paths of work.

%===================== system model ======================

\section{System Model and Main Results}
\label{sec:main}

Consider the DI over a DTPC with K-symbol memory: $\mathcal{P}=\left(\mathcal{X},\mathcal{Y},W_{\mathcal{P}}(\cdot|\cdot),K\right)$ consisting of an input alphabet $\mathcal{X}\subset\mathbb{R}_0^+$, an output alphabet $\mathcal{Y}\subset\mathbb{N}_0$, a Poissonian probability mass function $W_{\mathcal{P}}$ on $\mathcal{Y}$, and a channel memory $K\in\mathbb{N}_0^+$.
%\pau{Isn't it a bit weird to define the tuple $\mathcal{P}$ with the pmf which, at the end of the day, is exactly a Poisson distribution? Isn't the important parameter here the dark current that will define the (poissonian) pmf? I mean, it is actually not true that a DTPC is defined by the tuple claimed above, you additionally need that the pmf comes from a Poisson distribution. I would probably just avoid defining the tuple at all, and start describing the channel itself, adding the discrete time after, and finally add the memory. What do you think?}

%\yz{I think it is somehow a tradition in classical information theory to define a channel as a tuple of input alphabet, probability distribution, and output alphabet. But you are right we need to clarify the probability distribution is poissonian here.}
%\cd{I would leave the tuple but would additionally the definition of $W$ here}\pau{Cool!}
We divide the total duration of the process in $n+K$ time slots of duration $T_s$.
For all $t\in[1,n]$, the transmitter releases molecules with a rate $x_t$ modeled by the random variable (RV) $X_t$.
%(hence, we are considering a randomized release model). 
The transmission ends then, so during the rest of the protocol, for $t\in[n+1,n+K]$, $X_{t>n}\sim\delta_0$, with $\delta_0$ the Dirac delta centered at 0. In other words $x_{t>n}=0$.
The random variable $Y_t$ models the rate of absorbed molecules upon hitting the receiver at the $t$-th time slot, for all $t\in[1,n+K]$.

\begin{figure}[t]
    \centering
    \includegraphics[width=0.65\linewidth]{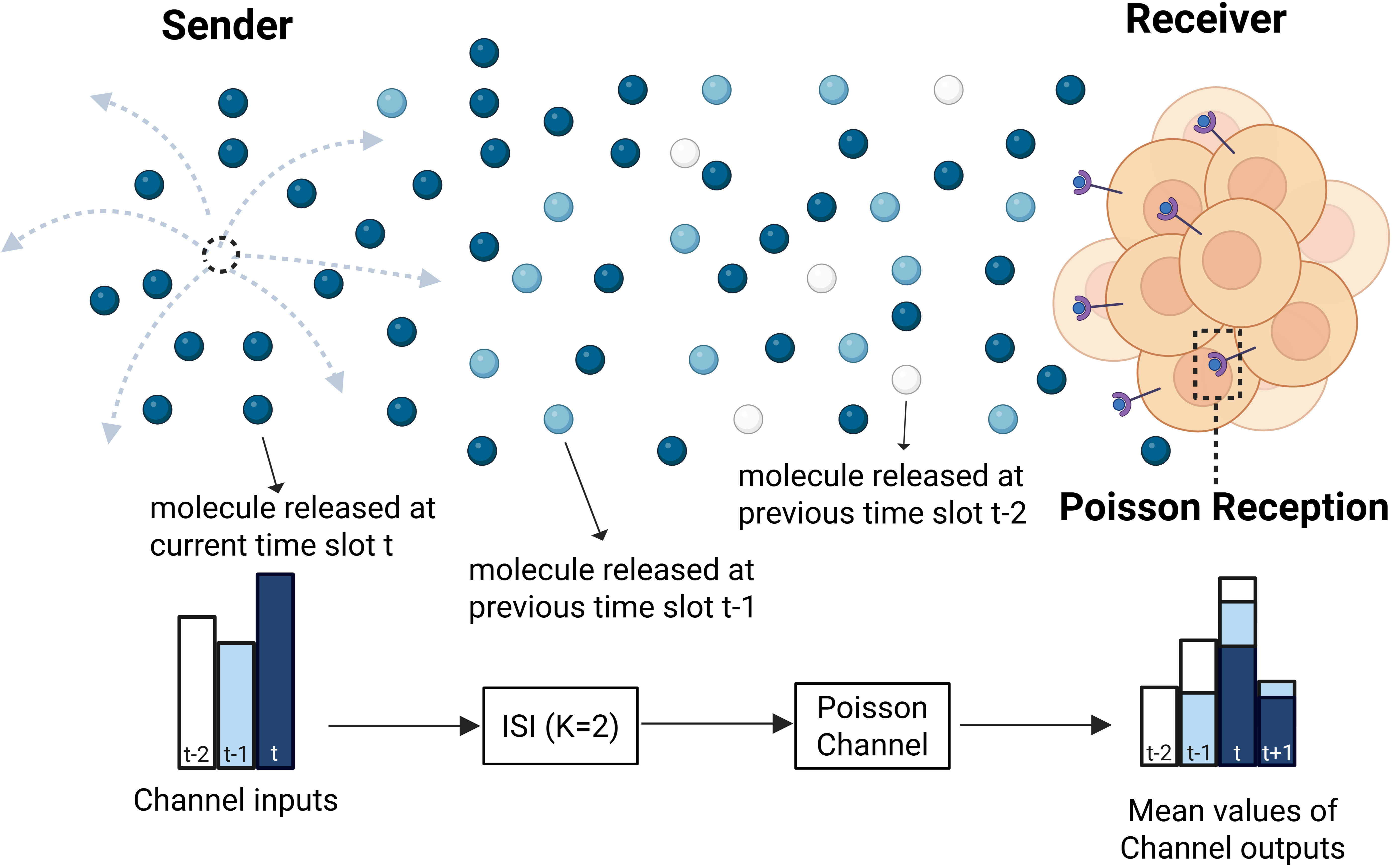}
    \caption{DTPC with ISI for MC, with parameters $K=2$ and $[p_0,p_1,p_2]=[0.6,0.3,0.1]$}
    \label{fig:ISI}
\end{figure}

The channel memory is modeled by a vector-valued probability sequence $\boldsymbol{p}=[p_0,\cdots,p_{K}]$, where each $p_k\in[0,1]$ denotes the probability that a molecule released at the current time slot $t$ hits the receiver in the next $k$-th period (for $k\in[K]$), as illustrated in Fig. \ref{fig:ISI}. We consider fully absorbing receivers ($\sum_{k=0}^{K}p_k=1$), meaning that at time $t+K$ all the molecules released at $t$ have been read \cite{gohari2016information,ferrari2022channel}. 
%\pau{Shouldn't we add the probability $p_K$ to avoid sub and super-indexes?}
At the $t$-th time slot, three sources contribute to the received molecules:
\begin{enumerate}
    \item Molecules released in the current time slot: $p_0x_tT_s$;
    \item Residue molecules released in earlier slots, known as the interference signal: $\sum_{k=1}^{K}p_kx_{t-k}T_s$;
    \item Noise molecules representing the non-ideality of the detector, known as dark current, and denoted as a nonnegative constant $\lambda_0$.
\end{enumerate}
%\pau{The indexes in the sum in 2. where $\sum_{t=1}^{K-1}$, I think they should be the ones I added, right? Also in Eq.~\eqref{eq:memory}, they were $\prod_{t=1}^{n+K}$.}

\begin{figure}[H]
    \centering
    \scalebox{1}{
\tikzstyle{farbverlauf} = [ top color=white, bottom color=white]
\tikzstyle{block} = [draw,top color=white, bottom color=white, rectangle, minimum height=2em, minimum width=2.5em]
\tikzstyle{input} = [coordinate]
\tikzstyle{sum} = [draw, circle,inner sep=0pt, minimum size=2mm,  thick]
\scalebox{1}{
\tikzstyle{arrow}=[draw,->] %{Latex[length=3mm]},
\begin{tikzpicture}[auto, node distance=2cm,>=latex']
\node[] (M) {};
\node[block,right=1cm of M] (enc) {Encoder};
\node[block, right=1.5cm of enc](ISI){ISI};
\node[block, right=.8cm of ISI] (channel) {DTPC};

\node[block, right=1cm of channel] (dec) {Decoder};
\node[below=.3cm of dec](i'){$i'\in[N]$};
\node[right=1.2cm of dec] (Mhat) {};
\node[input,right=.5cm of channel] (t1) {};
\node[input,above=1cm of t1] (t2) {};
\draw[-{Latex[length=1.5mm, width=1.5mm]},thick] (M) -- node[above]{$i\!\in\! [N]$} (enc);
\draw[-{Latex[length=1.5mm, width=1.5mm]},thick] (enc) --node[above]{ $X_{t-K}^t$} (ISI);
\draw[-{Latex[length=1.5mm, width=1.5mm]},thick](ISI)--node[above]{$X^{\star}_t$}(channel);
\draw[-{Latex[length=1.5mm, width=1.5mm]},thick] (channel) --node[above]{$Y_t$} (dec);
\draw[-{Latex[length=1.5mm, width=1.5mm]},thick] (dec) --node[above]{$i=i'?$} (Mhat);
\draw[-{Latex[length=1.5mm, width=1.5mm]},thick](i') -- (dec);
%\draw[-] (t1) -- (t2);
%\draw[->] (t2) -| (enc);
%\draw[->] (state) -| node[below]{ $S^n$} (dec);
\end{tikzpicture}}
}
    \caption{DI over DTPC with ISI}
    \label{fig:system model}
\end{figure}
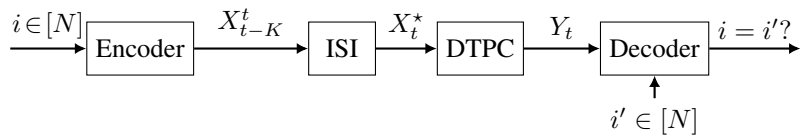

%\yz{I think $t$ should begin with $1$ to denote the index of time slots, but $k$ should begin with $0$ to denote the memory index with respect to the current time index. That is, Eq. ~\eqref{eq:memory} should be $\prod_{t=1}^{n+K}$, but in 2) should be $\sum_{k=1}^{K}$.}
%\cd{Yaning it correct.}\pau{Cool, I will take care of this!}
These many different sources of information motivate us to consider a Poisson reception process, which conveniently captures the statistical properties of molecule arrivals. Let $\boldsymbol{x}_{t-K}^t=(x_{t-K},\cdots,x_t)$ denote the $K+1$ most recent input symbols (release rates); and $X^{\star}_t=\sum_{k=0}^{K}p_kX_{t-k}T_s$, denote the RV representing the molecules absorbed at each time slot (see Figure \ref{fig:system model}). Again, we impose that the RVs with negative sub-indexes, which can be encountered in the regime $t< K$, are $X_{<0}\sim\delta_0$. That is, the samples $x_{<0}=0.$
% That is, the corresponding samples are always $x_{<0}=0$.

Following the model, given a channel input $X_t$, the channel output $Y_t$ is characterized by $Y_t\sim\text{Pois}(X_t^{\star}+\lambda_0)$, i.e.,
\begin{align}
    W_{\mathcal{P}}(y_t|\boldsymbol{x}_{t-K}^t)=\frac{e^{-(x_t^{\star}+\lambda_0)}(x_t^{\star}+\lambda_0)^{y_t}}{y_t!}.\nonumber
\end{align}

Additionally, as we consider fully absorbing receivers, we assume that given the $K$ previous input symbols $\boldsymbol{x}_{t-K}^t$ the channel transitions are independent, i.e.,
\begin{align}
    W_{\mathcal{P}}^{n+K}(\boldsymbol{y}^{n+K}|\boldsymbol{x}^n)
    &=\prod_{t=1}^{n+K}W_{\mathcal{P}}(y_t|\boldsymbol{x}_{t-K}^t)\nonumber\\
    &=\prod_{t=1}^{n+K}\frac{e^{-(x_t^{\star}+\lambda_0)}(x_t^{\star}+\lambda_0)^{y_t}}{y_t!}\label{eq:memory}.
\end{align}
%\pau{To be absolutely rigorous we should define an extended time frame $[t']=[t]\cup [K]$ to account for the $K$ time slots corresponding to the residual absorption when we have already ended the release of molecules. Of course, this has no effect in the proof, as we look at the limit. But it is not accurate to define $t\in[1,n]$ (beginning of the second paragraph in section II, and after that, in Eq.~\eqref{eq:memory} use values of $t>n$ in the sum. Maybe, easier than a new time frame, it would be better to define $t\in[1,n+K]$ directly, and then note that the random variables modeling the release of molecules $X_t$ and $Y_t$ are 0 for $t>n$, similarly to what I did for the $X^\star$. I think the second option is actually quite simple to understand for readers. If you agree, I will quickly modify it!} 
%\yz{I totally agree! It is easier to remove the part after $t>n$.}
%\cd{I agree also!}
%\pau{Great! I will change this.}
%\pau{We have to do something about how we define the sequences, it is not completely correct, though I don't think it affects the results. for $t\geq K$ and $t\leq n$ the definitions are correct. But not for the other (much smaller) regimes. Indeed, for $t< K$, $X^\star_t$ is not well defined (the random variable $X_{<0}$ has negative subindex), we can easily fix this one by saying that any $x\sim X_{<0}=0$.} 
Finally, we impose peak and average power constraints:
\begin{equation}\label{eq:power_constraints}
    x_t\leq \hat{E},\hspace{3pt} \forall t\in[1,n], \quad\text{and}\quad
    \sum_{t=1}^nx_t\leq n\bar{E},
\end{equation}
where $\hat{E}$ and $\bar{E}$ are two constants. Note that the constraints above can always be trivially extended to the full duration of the process; because, as discussed, in the regime $t\in[n+1,n+K]$, the input release rates are always 0. In the following, we define a DI code for DTPC with ISI.
\begin{definition}
    Let $\lambda_1,\lambda_2\geq0$, and $\lambda_1+\lambda_2<1$. Then, an $(n,N,K,\lambda_1,\lambda_2)$ DI code for DTPC $\mathcal{P}$ with ISI under peak and average molecule release rate constraints is a family $\left\{(u_i,\mathcal{D}_i)|i=1,\cdots,N\right\}$ of codewords $u_i\in\mathcal{X}^n$ satisfying $0\leq u_{i,t}\leq \hat{E}$ and $\sum_{t=1}^n u_{i,t}\leq n\bar{E}$ for all $t\in[1,n]$, and decoding regions $\mathcal{D}_i\subset{\mathbb{N}_0^+}^{n+K}$; such that the Type I error and the Type II error satisfy
    \begin{align}
        P_{e,1}(i)&\triangleq W_{\mathcal{P}}^{n+K}(\mathcal{D}^c_i|u_i)\leq \lambda_1,\quad \forall i\in[N], \label{eq:Pe1}\\
        P_{e,2}(i,{i'})&\triangleq W_{\mathcal{P}}^{n+K}(\mathcal{D}_{{i'}}|u_i)\leq \lambda_2,\quad \forall i\neq{i'}\in[N]. \label{eq:Pe2}
    \end{align}
\end{definition}
Each code can be characterized by the rate, suitably defined for DI in the superexponential scale as $R=\frac{\log N}{n\log n}$. A rate is said to be \emph{achievable}, if there exists an $(n,N,K,\lambda_1,\lambda_2)$ DI code that can attain it. Given a DTPC $\mathcal{P}$, we define the superexponential capacity $C_\text{ID}^\text{d}(\mathcal{P})$ as the supremum of all achievable rates in the limit of infinite channel uses:
\begin{equation}\label{eq:capacity}
C_\text{ID}^\text{d}(\mathcal{P})=\inf_{\lambda_1,\lambda_2>0}\liminf_{n\rightarrow\infty}\frac{1}{n\log n}\log N_{\text{ID}}^\text{d}(K,\lambda_1,\lambda_2),
\end{equation}
where, $N_{\text{ID}}^\text{d}(K,\lambda_1,\lambda_2)$ is the maximum number of words in the code. In this work we prove the following:
%The following theorem characterized the DI capacity of DTPC with ISI $\mathcal{P}$ under the peak power constraint $\hat{E}$ and the average power constraint $\bar{E}$.
%\yz{I added the peak power constraint and consider the minimum of $n\hat{E}$ and $n\bar{E}$ as the radius of the large hypersphere. Please check whether it makes scene or not.}
\begin{theorem}\label{thm:main1}
Assume that the number of ISI scales sub-linearly with code length $n$, i.e., $K=\lfloor2^{\kappa\log{n}}\rfloor$, then the DI capacity of a DTPC with ISI $\mathcal{P}$ under peak power constraint $x_t\leq \hat{E}$ and average power constraint $\sum_{t=1}^nx_t\leq nE$ is bounded by
    \begin{align}
        \frac{1-\kappa}{4}\leq C^\text{d}_\text{ID}(\mathcal{P})\leq \frac{1+\kappa}{2}.\nonumber
    \end{align}
\end{theorem}

We extend this result to a model with a noiseless strict-causal feedback link. 
In this model, we have for each message $i\in[N]$ a vector-valued function
\(
\boldsymbol{f}_{i}=\left[f_{i}^1,\cdots,f_{i}^n\right],
\)
with $f_{i}^t:\mathcal{Y}^{t-1}\mapsto \mathcal{X}$, the output of which will be the input. Notice that the first function $f_{i}^1$ can only have 0 as argument (because there is no output at $t=0$), and that there is no point in modeling the feedback for $f_i^{t>n}$, because, as there will be no more releases, it is useless in this regime. We denote the set of all possible $\boldsymbol{f}_{i}$ as $\mathcal{F}_n$. 
%\pau{Is the feedback removed for $t\in(n,n+K]$?} \yz{We can define $t\in[n,n+K]$ as zeros as you suggest before.} \pau{Yes! I will handle this.}
%In the following, we define DIF code for DTPC with ISI.
\begin{definition}
    Let $\lambda_1,\lambda_2\geq0$, and $\lambda_1+\lambda_2<1$. Then, an $(n,N,K,\lambda_1,\lambda_2)$ DIF code for DTPC with ISI $\mathcal{P}$ under peak and average molecule release rate constraints is a family $\left\{(\boldsymbol{f}_i,\mathcal{D}_i)|i\in[N]\right\}$, with feedback encoding functions $\boldsymbol{f}_i\in\mathcal{F}_{n}$ serving as codewords and satisfying $0<f_i^t(y^{t-1})\leq \hat{E}$ for all $t\in\left[1,n\right]$ and $\frac{1}{n}\sum_{t=1}^{n}{f^t_i}(y^{t-1})\leq \bar{E}$, and decoding regions $\mathcal{D}_i\subset \mathbb{N}_0^{n+K}$; such that the Type I error and the Type II error are bounded as follows:
    \begin{align}
        P_{e,1}(i)&=W_{\mathcal{P}}^{n+K}\left(\mathcal{D}_i^c|\boldsymbol{f}_i\right)\leq \lambda_1,\quad \forall i\in[N],\\
        P_{e,2}(i,{i'})&=W_{\mathcal{P}}^{n+K}\left(\mathcal{D}_{{i'}}|\boldsymbol{f}_i\right)\leq \lambda_2, \quad \forall i\neq {i'}\in[N].
    \end{align}
\end{definition}

\begin{figure}[H]
    \centering
    \scalebox{1}{
\tikzstyle{farbverlauf} = [ top color=white, bottom color=white]
\tikzstyle{block} = [draw,top color=white, bottom color=white, rectangle,
minimum height=2em, minimum width=2.5em]
\tikzstyle{input} = [coordinate]
\tikzstyle{sum} = [draw, circle,inner sep=0pt, minimum size=2mm,  thick]
\scalebox{1}{
\tikzstyle{arrow}=[draw,->] %{Latex[length=3mm]},
\begin{tikzpicture}[auto, node distance=2cm,>=latex']
\node[] (M) {};
\node[block,right=1cm of M] (enc) {Encoder};
\node[block, right=2.5cm of enc](ISI){ISI};
\node[block, right=.8cm of ISI] (channel) {DTPC};
\node[input,right=.8cm of channel](p1){};
\node[input,above=1.1cm of p1](p2){};
\node[input,above=.75cm of enc](p3){};
\node[block, right=1cm of channel] (dec) {Decoder};
\node[below=.3cm of dec](i'){$i'\in[N]$};
\node[right=1.2cm of dec] (Mhat) {};
\node[input,right=.5cm of channel] (t1) {};
\node[input,above=1cm of t1] (t2) {};
\draw[-{Latex[length=1.5mm, width=1.5mm]},thick] (M) -- node[above]{$i\!\in\![N]$} (enc);
\draw[-{Latex[length=1.5mm, width=1.5mm]},thick] (enc) --node[above]{ $X_{t}=f_i^t(Y^{t-1})$} (ISI);
\draw[-{Latex[length=1.5mm, width=1.5mm]},thick](ISI)--node[above]{$X^{\star}_t$}(channel);
\draw[-{Latex[length=1.5mm, width=1.5mm]},thick] (channel) --node[above]{$Y_t$} (dec);
\draw[-{Latex[length=1.5mm, width=1.5mm]},thick] (dec) --node[above]{$i=i'?$} (Mhat);
\draw[-{Latex[length=1.5mm, width=1.5mm]},thick](i') -- (dec);
\draw[-,thick](p1) -- (p2);
\draw[-,thick](p2) -- (p3);
\draw[-{Latex[length=1.5mm, width=1.5mm]},thick](p3) -- (enc);
%\draw[-] (t1) -- (t2);
%\draw[->] (t2) -| (enc);
%\draw[->] (state) -| node[below]{ $S^n$} (dec);
\end{tikzpicture}}
}
    \caption{DIF over DTPC with ISI}
    \label{fig:system IDF}
\end{figure}
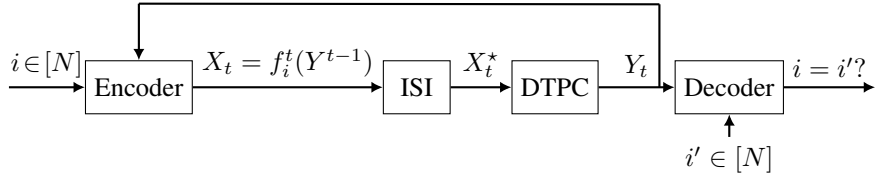
As we will see, the introduction of noiseless feedback elevates the code size growth scaling from super-exponential to double-exponential. The rates for DIF are therefore suitable defined in this regime as $R=\frac{\log\log{N}}{n}$, and the corresponding double exponential capacity $C_\text{IDF}^\text{d}(\mathcal{P})$ is defined through its supremum:
\begin{equation}\label{eq:double_capacity}
C_\text{IDF}^\text{d}(\mathcal{P})=\inf_{\lambda_1,\lambda_2>0}\liminf_{n\rightarrow\infty}\frac{1}{n}\log\log N_{\text{IDF}}^\text{d}(K,\lambda_1,\lambda_2),
\end{equation}
where $N_\text{IDF}^\text{d}(K,\lambda_1,\lambda_2)$ is the maximum number of elements in an $(n,N,K,\lambda_0,\lambda_1)$ IDF code. In this work, we lower bound this capacity as follows:
\begin{theorem}\label{thm:main2}
%\pau{The minimization over the constraints is missing.} \yz{I am thinking about how we can use the average power constraint in the encoding scheme we provide later for the proof, because under average power constraint we can transmit with power of $K(n,\kappa)\bar{E}$ instead of $\hat{E}$, which goes to infinity when $n$ goes to infinity. I mean, in any case we may always only need to consider peak power constraint.} 
The DIF capacity of a DTPC with ISI $\mathcal{P}$ under peak power constraint $x_t\leq \hat{E}$
%average power constraint $\sum_{t=1}^n x_t \leq n\bar{E}$ 
is lower-bounded by
\begin{align}
        C^d_{IDF}(\mathcal{P})
        &\geq \lim_{n\to\infty}\frac{1}{K+1}\sum_{k=0}^{K}H(\text{Pois}(p_k\hat{E}T_s+\lambda_0))\nonumber\\
        &\approx \frac{1}{2}\log{\left(2\pi e \hat{E}T_s\right)}.
    \end{align}
\end{theorem}
\medskip

%========================== proof ===================================
\section{Proof of Main Results}
\label{sec:proof}
% \subsection{Technical tools for Theorem \ref{thm:main1}}

\subsection{Proof of Theorem \ref{thm:main1}}
%\pau{The Euclidean distance should also be defined (as it is used further down the proof). It might be better to define directly the p-Schatten norm, so that both L1 and L2 norms (needed for the TVD and Euclidean distance respectively) are included in a single definition.}
%\yz{I will take care of these definitions.}
Before proving Theorem \ref{thm:main1}, we need to define some distance measures. Let $Q_1$ and $Q_2$ be two probability distributions defined on a countable measurable space $\mathcal{X}$, then the $\mathcal{L}_1$-norm, defined as
\begin{align}
    \norm{Q_1-Q_2}_1\triangleq\sum_{a\in\mathcal{X}}\big|Q_1(a)-Q_2(a)\big|
\end{align}
can be used to calculate the total variation distance:
\begin{align}
\!\!\!\!\delta(Q_1,Q_2)\triangleq\max_{\mathcal{A}\subseteq \mathcal{X}}\big|Q_1(\mathcal{A})-Q_2(\mathcal{A})\big|=\frac{1}{2}\norm{Q_1-Q_2}_1\label{eq:TVD&norm1},
\end{align}
which is related to the Bhattacharyya coefficient 
\(
F(Q_1,Q_2)=\sum_{a\in\mathcal{X}}\sqrt{Q_1(a)Q_2(a)},
\)
as follows
\begin{equation}
1-F(Q_1,Q_2)\leq \delta(Q_1,Q_2)\leq \sqrt{1-F(Q_1,Q_2)^2}\label{eq:TVD&F}.
\end{equation}
Finally, the Euclidean distance (the $\mathcal{L}_2$-norm) between two $n$-length vectors $v^n$ and $w^n$ is given, as usual, by
\begin{align}
    \norm{v^n-w^n}_2=\sqrt{\sum_{t=1}^n(v_t-w_t)^2}.\nonumber
\end{align}

The achievability part: $C_\text{ID}^\text{d}(\mathcal{P})\geq \frac{1-\kappa}{4}$, was proved in \cite{dIDMCISI}. Only the converse $C_\text{ID}^\text{d}(\mathcal{P})\leq \frac{1+\kappa}{2}$ is needed. By the definition of DI code, the output distributions of channel output with distinct codewords $u_i=x^n$ and $u_{{i'}}=x'^n$ must have 
\begin{align}
    \delta[{W_{\mathcal{P}}(\cdot|x^n),W_{\mathcal{P}}(\cdot|x'^n)}]
    &\overset{(a)}{=}\frac{1}{2}\norm{W_{\mathcal{P}}(\cdot|x^n)-W_{\mathcal{P}}(\cdot|x'^n)}_1\nonumber\\
    &\overset{(b)}{\geq} 1-\lambda_1-\lambda_2\triangleq\delta,\nonumber
\end{align}
where $(a)$ follows from Eq.~\eqref{eq:TVD&norm1}, and $(b)$ from Eqs.~\eqref{eq:Pe1} and~\eqref{eq:Pe2}.
%\pau{Watch out for the sum indexes (I will take care of this).}
Let us now choose $r>0$, such that
\begin{align}
    (2r)^2
    &\triangleq -\ln{(1-\delta^2)}\nonumber\\
    &\overset{(a)}{\leq} -\ln\left[F\left(W_{\mathcal{P}}^{n+K}(\cdot|x^n),W_{\mathcal{P}}^{n+K}(\cdot|x'^n)\right)^2\right]\nonumber\\
    &\overset{(b)}{=}\sum_{t=1}^{n+K}-\ln{\left[F\left(W_{\mathcal{P}}\left(y_t|{\boldsymbol{x}}_{t-K}^{t}\right), W_{\mathcal{P}}\left(\cdot|{\boldsymbol{x}'}_{t-K}^{t}\right)\right)^2\right]},
    \label{eq:r}
\end{align}
where $(a)$ follows from Eq.~\eqref{eq:TVD&F}, and $(b)$ follows from Eq.~\eqref{eq:memory}.
We can expand the Bhattacharyya coefficient between these Poisson distributions inside the logarithm above as:
\begin{align}
    F&\left[ W_{\mathcal{P}}\left(\cdot|\boldsymbol{x}_{t-K}^{t}\right),W_{\mathcal{P}}\left(\cdot|\boldsymbol{x}'^{t}_{t-K}\right)\right]^2\nonumber\\
    &=\left[\sum_{y\in\mathbb{N}_0}\sqrt{W_{\mathcal{P}}\left(y|\boldsymbol{x}_{t-K}^{t}\right)W_{\mathcal{P}}\left(y|\boldsymbol{x}'^{t}_{t-K}\right)}\right]^2\nonumber\\
    %&=\left[\sum_{y_t\in\mathbb{N}_0}\sqrt{\frac{e^{-(x_t^{\star}+\lambda_0)}(x_t^{\star}+\lambda_0)^{y_t}}{y_t!}}\sqrt{\frac{e^{-({x'}_t^{\star}+\lambda_0)}({x'}_t^{\star}+\lambda_0)^{y_t}}{y_t!}}\right]^2\nonumber\\
    &=\left[e^{-\left(\frac{x_t^{\star}+{x'}_t^{\star}}{2}+\lambda_0\right)}\cdot\sum_{y_t\in\mathbb{N}_0}\frac{\sqrt{(x_t^{\star}+\lambda_0)({x'}_t^{\star}+\lambda_0)}^{y_t}}{y_t!}\right]^2\nonumber\\
    &=e^{-\left(x_t^{\star}+{x'}_t^{\star}+2\lambda_0\right)}\cdot e^{2\sqrt{(x_t^{\star}+\lambda_0)({x'}_t^{\star}+\lambda_0)}}\nonumber\\
    &=e^{-\left(\sqrt{x_t^{\star}+\lambda_0}-\sqrt{{x'}_t^{\star}+\lambda_0}\right)^2}.
    \label{eq:F}
\end{align}
Combining Eqs.~\eqref{eq:r} and \eqref{eq:F}, it is clear that
\begin{align}
    (2r)^2\leq \sum_{t=1}^n\left(\sqrt{x_t^{\star}+\lambda_0}-\sqrt{{x'}_t^{\star}+\lambda_0}\right)^2\nonumber.
\end{align}
So, reparameterizing $s_t=\sqrt{x_t^{\star}+\lambda_0}$ we reach a minimum Euclidean distance condition that any two (reparametrized) codewords $s^n$ and $s'^n$ have to fulfill:
\begin{align}
    2r\leq  \sqrt{\sum_{t=1}^n(s_t-s'_t)^2}=\norm{s^n-s'^n}_2\label{eq:Eucl_converse}.
\end{align}
We have to redefine the power constraints for these new codewords. For the average release rate constraint we observe:
\begin{align*}
\sum_{t=1}^n s_t^2&=n\lambda_0+\!\sum_{t=1}^n\sum_{k=0}^K p_kx_{t-k}T_s\\
&\leq n\lambda_0+\!\sum_{t=1}^n K(n,\kappa) x_{t}T_s\\
&\leq n\lambda_0+n\bar{E}K(n,\kappa)T_s\triangleq \bar{l}^2,
\end{align*}
where the first inequality follows from the fact that $p_k x_{t-k}\leq x_t$, and the second from the power constrain in the original code book.
Similarly, for the peak molecule release rate constraint, we have for all $t\in[1,n]$:
\begin{align*}
s_t^2=x_t^{\star}+\lambda_0=\lambda_0+\sum_{k=1}^K p_k x_{t-k}T_s&\leq\lambda_0+K(n,\kappa)x_t T_s\\
&\leq\lambda_0+\hat{E}K(n,\kappa) T_s.
\end{align*}
Notice that we can trivially sum this second constraint for each $t$ to get $\sum_{t=1}^n s_t^2=n\lambda_0+n\hat{E}K(n,\kappa) T_s\triangleq\hat{l}$. In this form, it is clear that the code words can only fulfill both constraints if they lie simultaneously inside hyperspheres of radii $\bar{l}$ and $\hat{l}$, so we might as well take the smaller one. Indeed, let us define $E=\min\{\bar{E},\hat{E}\}$, then this minimal radius is $l=\sqrt{n\lambda_0+nEK(n,\kappa)T_s}$.
\begin{figure}[H]
    \centering
    \definecolor{lightblue}{rgb}{0.690, 0.886, 0.910}
\scalebox{0.85}{

\begin{tikzpicture}[scale=.55][thick]

\draw[thick] (0,0) circle (4cm);
\draw[thick,dashed] (0,0) circle (5cm);
%% Entire Spheres
\draw (0,0) circle (1cm);
\draw [fill=lightblue, fill opacity=0.4] (0,0) circle (1cm);

\draw (2,0) circle (1cm);
\draw [fill=lightblue, fill opacity=0.4] (2,0) circle (1cm);

\draw (1,1.73) circle (1cm);
\draw [fill=lightblue, fill opacity=0.4] (1,1.73) circle (1cm);

\draw (-1,1.73) circle (1cm);
\draw [fill=lightblue, fill opacity=0.4] (-1,1.73) circle (1cm);

\draw (-2,0) circle (1cm);
\draw [fill=lightblue, fill opacity=0.4] (-2,0) circle (1cm);

\draw (-1,-1.73) circle (1cm);
\draw [fill=lightblue, fill opacity=0.4] (-1,-1.73) circle (1cm);

\draw (1,-1.73) circle (1cm);
\draw [fill=lightblue, fill opacity=0.4] (1,-1.73) circle (1cm);

%% Partial Spheres

\draw (3,-1.73) circle (1cm);
\draw [fill=lightblue, fill opacity=0.4] (3,-1.73) circle (1cm);

% \node at (3,1.73) {$.$};
\draw (3,1.73) circle (1cm);
\draw [fill=lightblue, fill opacity=0.4] (3,1.73) circle (1cm);

\draw (0,2*1.73) circle (1cm);
\draw [fill=lightblue, fill opacity=0.4] (0,2*1.73) circle (1cm);

\draw (-3,1.73) circle (1cm);
\draw [fill=lightblue, fill opacity=0.4] (-3,1.73) circle (1cm);

\draw (-3,-1.73) circle (1cm);
\draw [fill=lightblue, fill opacity=0.4] (-3,-1.73) circle (1cm);

\draw (0,-2*1.73) circle (1cm);
\draw [fill=lightblue, fill opacity=0.4] (0,-2*1.73) circle (1cm);

%% Arrow
\draw (0,0)[] -- (4,0) node [right,font=\small] {$l$};

\draw (-3,-1.73)[]-- (-3.707,-2.437)    node [below,font=\small] {$r$};

\draw (0,0)[dashed]--(0,5) node [above, font=\small] {$l+r$};
\end{tikzpicture}}
    \caption{Packing of many $S_{u_i}(n,r)$ spheres into the bigger $S_0(n,l)$, and the extended $S_0(n,l+r)$ containing them all.}
    \label{fig:packing}
\end{figure}
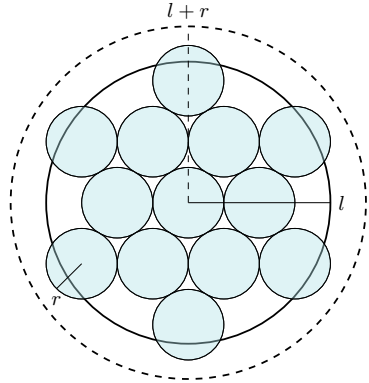

Now, we have seen in Eq.~\ref{eq:Eucl_converse} that all different code words have to be at the minimum Euclidean distance $2r$. This is the same as saying that we can only have a single code word in a hypersphere $S_u{_i}(n,r)$ of radius $r$. Due to the power constraints, all these words (sphere centers) have to lie inside a bigger hypersphere $S_0(n,l)$ of radius $l$. The problem finding the maximum number of words in our code is therefore reduced to the geometrical problem of counting the number $N$ of $S_u{_i}(n,r)$ that can be packed in $S_0(n,l)$, see Fig. \ref{fig:packing}. This number can clearly be upper-bounded as follows: 
\begin{align*}
    N\leq\frac{\text{Vol}[S_{0}(n,l+r)]}{\text{Vol}[S_{u_i}(n,r)]}= \frac{(l+r)^n}{r^n}\leq\frac{(2l)^n}{r^n}
\end{align*}
where we have applied the volume of a hypersphere $\text{Vol}[S_{u_i}(n,r)]={\pi^{\frac{n}{2}}r^n}/{\Gamma\left(\frac{n}{2}+1\right)}$ and the last inequality follows from $r\leq l$. Now, imposing that $K(n,\kappa)=\lfloor2^{\kappa\log n}\rfloor$, 
%(see the statement of the theorem)
we can calculate the logarithm:
\begin{align*}
    \log{N}&\leq n\left[1+\log l +\log r\right]\\
    &\leq n\left[\frac{1}{2}\log n +\frac12\log\left(\lambda_0+ET_s2^{\kappa\log n}\right)+\mathcal{O}(1)\right]\\
    &=n\log{n}\cdot\frac{1+\kappa}{2}+\mathcal{O}(n).
\end{align*}
Inserting this into Eq.~\eqref{eq:capacity} we obtain the claimed capacity upper bound, completing the proof of Theorem \ref{thm:main1}.\hfill\qed

\subsection{Proof of Theorem \ref{thm:main2}}

We extend the encoding scheme for DIF over point-to-point DMCs, outlined in \cite{IDF}, to our DIF problem over a DTPC with ISI. There $n$ uses of the channel are first used to create uniform common randomness between the two parties through the feedback link. The randomness is then used to agree on a particular code book. Then, $\sqrt{n}$ uses of the channel are used to identify a code word in this code book. We have to proof that a double exponential amount of code words (scattered through the different code books) can be reliably identified. 

We start from an $(m,N,K,\lambda_1,\lambda_2)$ DIF code, where $m=n+\lceil\sqrt{n}\rceil$. The encoder starts using $n$ times the channel to send $(K+1)\lceil\!\frac{n}{K+1}\!\rceil$ times the sequence $\boldsymbol{x}^{\star}\triangleq(\hat{E},0,\cdots,0)$ of length $(K+1)$, as illustrated in Table \ref{table:encdoing}. We denote the corresponding $(K+1)$-length output sequences $\boldsymbol{s_i}$, which can vary due to the stochastic nature of the channel and the ISI.

\begin{center}
\begin{table}[H]
\centering
    \renewcommand{\arraystretch}{1.5} 
        \caption{Feedback scheme and the corresponding outputs}
    \label{table:encdoing}
    \setlength{\tabcolsep}{6pt}
    \scalebox{0.9}{
    \begin{tabular}{|c|c|c|c|c|c|c|}
        \hline
        Time & $j(K+1)+1$ & $j(K+1)+2$ & $\cdots$ & $(j+1)(K+1)$ \\
        \hline
        Input & $\hat{E}$ & $0$ & $\cdots$ & $0$ \\
        \hline
        Receptions & $y_1^j=p_0\hat{E}T_s$ & $y_2^j=p_1\hat{E}T_s$ & $\cdots$ & $y_{K+1}^j=p_{K}\hat{E}T_s$\\
        \hline
        Output & \multicolumn{4}{c|}{$\boldsymbol{s_j}=(y_1^j,y_2^j,\dots,y_{K+1}^j)$} \\
        \hline
    \end{tabular}
    }

    \vspace{-3em}
\end{table}
\end{center}

The received sequence $\boldsymbol{s}^{\lceil\!\frac{n}{K+1}\!\rceil}=(\boldsymbol{s}_1,\cdots,\boldsymbol{s}_{\lceil\!\frac{n}{K+1}\!\rceil})=y^n$ is then shared to the encoder via the noiseless feedback link. This is already a source of shared randomness, but for coding reasons we additionally want it to be uniform. The price to pay to get an almost-uniform source from the current construction is small if we use typicality. Namely, let
$\mathcal{D}^*=\mathcal{T}_{\epsilon}^{\lceil\!\frac{n}{K+1}\!\rceil}(P(\cdot|\boldsymbol{x}^{\star}))$
be the set of typical sequences conditioned to our fixed input $\boldsymbol{x^\star}$, then, for our channel 
$P(\cdot|\boldsymbol{x}^{*})=\prod_{k=0}^{K}\frac{e^{-\left(p_k\hat{E}T_s+\lambda_0\right)}\left(p_k\hat{E}T_s+\lambda_0\right)^{y_{k+1}}}{y_{k+1}!}$, and $\boldsymbol{s}^{\lceil\!\frac{n}{K+1}\!\rceil}\in\mathcal{D}^*$:
\begin{align}
\!\!\!P(\boldsymbol{s}^{\lceil\!\frac{n}{K+1}\!\rceil}\big|\boldsymbol{x}^{\star\lceil\!\frac{n}{K+1}\rceil})&\doteq 2^{-\lceil\frac{n}{K+1}\!\rceil\sum_{k=0}^{K}H(\text{Pois}(p_kT_s\hat{E}+\lambda_0))},\label{typical1}\\
        |\mathcal{D}^*|&\doteq 2^{\lceil\!\frac{n}{K+1}\!\rceil\sum_{k=0}^{K}H(\text{Pois}(p_kT_s\hat{E}+\lambda_0))},\label{typical2}\\
        P(\mathcal{D}^{*}|\boldsymbol{x}^{\lceil\!\frac{n}{K+1}\!\rceil})& \doteq1\label{typical3},  
\end{align}
where $\doteq$ denotes the asymptotic equivalence in $n$. The first equality \eqref{typical1} tells us that the sequences in $\mathcal{D}^*$ have asymptotically the same probability of appearing. Thus, we can convert the shared randomness source to a uniform one by allowing only typical sequences, with some error given by the set $e=\mathcal{S}^{\lceil\!\frac{n}{K+1}\!\rceil}\setminus \mathcal{D}^{\star}$.

Now, let $\left\{F_{i}:\mathcal{D}^* \mapsto [M]|i=1,\cdots,N\right\}$ be families of hashing functions. Each one mapping $M<N$ sequences in $\mathcal{D}^*$ to a hash $l\in[M]$ uniformly at random, that is, $Pr\left[F_i(\boldsymbol{s}^{\lceil\!\frac{n}{K+1}\!\rceil})=l\right]=\frac{1}{M}$. The mappings $F_{i}$, and most importantly their support, serve as prior knowledge for the encoder and the decoder. At this point, notice the similarities of this model with regular randomized identification.

So, if the message $i\in[N]$ has to be sent, the transmitter will calculate $F_{i}(\boldsymbol{s}^{\lceil\!\frac{n}{K+1}\!\rceil})$, and use the resulting hash $l$ as seed of an
$\left(\lceil\sqrt{n}\rceil,{M},2^{-\lceil\sqrt{n}\rceil\delta}\right)$
transmission code (notice that we are now using the remaining $\lceil\sqrt{n}\rceil$ uses of the channel to complete the total of $m$ claimed in the beginning).
$\mathcal{C}'=\left\{\left(\boldsymbol{c}({l}),\mathcal{D}'_{l}\right)\big|{l}\in[M]\right\}$.
Similarly, if the decoder wants to identify the message $i'$, he calculates $F_{i'}(\boldsymbol{s}^{\lceil\!\frac{n}{K+1}\!\rceil})$ and checks whether the decoded $\sqrt{n}$-sequence $\hat{F}_i=F_{i'}$ concluding the identification hypothesis test. 
If $\boldsymbol{s}^{\lceil\!\frac{n}{K+1}\!\rceil}\in e$, an error is declared. 
%$\boldsymbol{s}^{\lceil\!\frac{n}{K+1}\!\rceil}\in\mathcal{D}^*$ via feedback, the encoders calculate and transmit $F_{i}(\boldsymbol{s}^{\lceil\!\frac{n}{K+1}\!\rceil})$, using a standard $\left(\lceil\sqrt{n}\rceil,{M},2^{-\lceil\sqrt{n}\rceil\delta}\right)$ transmission code denoted as $\mathcal{C}'=\left\{\left(\boldsymbol{c}({l}),\mathcal{D}'_{l}\right)\big|{l}\in[M]\right\}$. Notably, we need to choose the code words under power constraints. If $\boldsymbol{s}^{\lceil\!\frac{n}{K+1}\!\rceil}\in e$, an error is declared. But by Eq.~\eqref{typical3}, this error occurs with a probability close to $0$. The decoder first computes $F_{i'}(\boldsymbol{s}^{\lceil\!\frac{n}{K+1}\!\rceil})$. Then, it decodes $\hat{{F}_{i}}$ based on $y_{(K+1)\lceil\!\frac{n}{K+1}\!\rceil+1}^m$. If ${F}_{i'}=\hat{F}_{i}$, it concludes $i=i'$; otherwise, $i\neq i'$.
Overall, our DIF code can be defined with
\begin{align*}
    \boldsymbol{f}_{i}(y^m) &= \left[\boldsymbol{x}^{*\lceil\!\frac{n}{K+1}\!\rceil},\boldsymbol{c}\left(F_i\left(\boldsymbol{s}^{\lceil\!\frac{n}{K+1}\!\rceil}\right)\right)\right],\nonumber\\
    \mathcal{D}_{i}&=\bigcup_{\boldsymbol{s}^{\lceil\!\frac{n}{K+1}\!\rceil}\in \mathcal{D}^*}\left\{\boldsymbol{s}^{\lceil\!\frac{n}{K+1}\!\rceil}\times \mathcal{D}'_{{F}_{{i}}\left(\boldsymbol{s}^{\lceil\!\frac{n}{K+1}\!\rceil}\right)}\right\},\quad \forall {i}\in{[N]}.
\end{align*}

For all ${i}\in{[N]}$, the type I error probability $P_{e,1}({i})$ can be upper-bounded by:
\begin{align*}
    W_{\mathcal{P}}^{m}(\mathcal{D}_{{i}}^c|\boldsymbol{f}_{{i}})&=\sum_{\boldsymbol{s}^{\lceil\!\frac{n}{K+1}\!\rceil}\in \mathcal{D}^*}P^{\lceil\!\frac{n}{K+1}\!\rceil}\left(s^{\lceil
     \frac{n}{K+1}\rceil}\big|\boldsymbol{x}^{*\lceil\!\frac{n}{K+1}\!\rceil}\right)\nonumber\\
     &\hspace{1.8cm}\cdot W_{\mathcal{P}}^{\lceil\sqrt{n}\rceil}\left(D'^c_{{F}_{{i}}\left(\boldsymbol{s}^{\lceil\!\frac{n}{K+1}\!\rceil}\right)}\bigg|\boldsymbol{c}\right)\\
     &\le 2^{-\lceil\sqrt{n}\rceil\delta}=\mathcal{O}(n).
\end{align*}
Thus, we can achieve an arbitrary small type I error probability with sufficient large code length.
For the type II error probability we separate the sum into the following subsets:
\begin{align*}
    F_{i}\cap F_{i'}&:= \left\{\boldsymbol{s}^{\lceil\!\frac{n}{K+1}\!\rceil}\in\mathcal{D}^*|F_{i}\left(\boldsymbol{s}^{\lceil\!\frac{n}{K+1}\!\rceil}\right)=F_{i'}\left(\boldsymbol{s}^{\lceil\!\frac{n}{K+1}\!\rceil}\right)\right\},\\
    F_{i}\setminus F_{i'}&:= \left\{\boldsymbol{s}^{\lceil\!\frac{n}{K+1}\!\rceil}\in\mathcal{D}^*|F_{i}\left(\boldsymbol{s}^{\lceil\!\frac{n}{K+1}\!\rceil}\right)\ne F_{i'}\left(\boldsymbol{s}^{\lceil\!\frac{n}{K+1}\!\rceil}\right)\right\}.
\end{align*}
Then, for each pair ${i},{{i'}}\in{[N]}$, where ${i}\neq{{i'}}$, $P_{e,2}({i},{{i'}})$ can be upper-bounded by:
\begin{align*}
    &W_{\mathcal{P}}^{m}(\mathcal{D}_{{i'}}|\boldsymbol{f}_{i})\\
    &\leq \sum_{\boldsymbol{s}^{\lceil\!\frac{n}{K+1}\!\rceil}\in \{F_{i}\cap F_{{i'}}\}} P^{\lceil\!\frac{n}{K+1}\!\rceil}\left(\boldsymbol{s}^{\lceil\!\frac{n}{K+1}\!\rceil}\big|\boldsymbol{x}^{*\lceil\!\frac{n}{K+1}\!\rceil}\right)\\
    &\quad +\sum_{\boldsymbol{s}^{\lceil\!\frac{n}{K+1}\!\rceil}\in \{F_{i}\setminus F_{{i'}}\}} P^{\lceil\!\frac{n}{K+1}\!\rceil}(\boldsymbol{s}^{\lceil\!\frac{n}{K+1}\!\rceil}|\boldsymbol{x}^{*\lceil\!\frac{n}{K+1}\!\rceil})\cdot 2^{-\lceil\sqrt{n}\rceil\delta}\\
    &\le\frac{|F_{i}\cap F_{{i'}}|}{|\mathcal{D}^*|}+\mathcal{O}(n).
\end{align*}
To bound the first term in the last inequality above, let us define an auxiliary RV $\Psi_{\boldsymbol{s}^{\lceil\!\frac{n}{K+1}\!\rceil}}$ for all $\boldsymbol{s}^{\lceil\!\frac{n}{K+1}\!\rceil}\in \mathcal{D}^*$ as:
\begin{align*}
    \Psi_{\boldsymbol{s}^{\lceil\!\frac{n}{K+1}\!\rceil}}(F_{{i'}})=\left\{
        \begin{array}{cc}
             1,&\boldsymbol{s}^{\lceil\!\frac{n}{K+1}\!\rceil}\in \{F_{i}\cap F_{{i'}}\}  \\
             0,&\boldsymbol{s}^{\lceil\!\frac{n}{K+1}\!\rceil}\in \{F_{i}\setminus F_{{i'}}\}
        \end{array},
    \right.
\end{align*}
with probability $Pr\left[\Psi_{\boldsymbol{s}^{\lceil\!\frac{n}{K+1}\!\rceil}}(F_{{i'}})=1\right]=\frac{1}{M}$. The following lemma bounds this object through an arbitrarily defined $\lambda_2$.
\begin{lemma}\cite{IDF}
    For $\lambda \in(0,1)$, and $E\left[\Psi_{\boldsymbol{s}^{\lceil\!\frac{n}{K+1}\!\rceil}}\right]=\frac{1}{M}<\lambda_2$,
    \begin{align*}
        Pr\left[\frac{1}{|\mathcal{D}^*|}\sum_{\boldsymbol{s}^{\lceil\!\frac{n}{K+1}\!\rceil}\in \mathcal{D}^*}\!\!\!\!\!\!\!\Psi_{\boldsymbol{s}^{\lceil\!\frac{n}{K+1}\!\rceil}}(F_{{i'}})\!>\!\lambda_2\right]
        \!\!<\!2^{-|\mathcal{D}^*| (\lambda_2\log(M)-1)}.
    \end{align*}
\end{lemma}
Then, for all pairs $(i,{i'})\in[N]^2$, $i\neq{i'}$, we have
\begin{align*}
    &1-Pr\left[ \bigcap_{{i'}\in[N],{i'}\ne i}\left\{\frac{1}{|\mathcal{D}^*|}\sum_{\boldsymbol{s}^{\lceil\!\frac{n}{K+1}\!\rceil}\in \mathcal{D}^*}\Psi_{\boldsymbol{s}^{\lceil\!\frac{n}{K+1}\!\rceil}}(F_{{i'}})\le\lambda_2\right\}\right]\\
    &\quad \leq  (N-1)\cdot 2^{-2^{\lceil\!\frac{n}{K+1}\!\rceil\sum_{k=0}^{K} H\left(\text{Pois}\left(p_k\hat{E}T_s+\lambda_0\right)\right)}\cdot(\lambda_2\log{M}-1)}
\end{align*}
which has to be smaller than $1$. Meaning that the maximum value of $N$ we can choose is:
\begin{align}
    N
    &=2^{2^{\frac{n}{K+1}\sum_{k=0}^{K}H(\text{Pois}(p_k\hat{E}T_s)+\lambda_0)}}.
\end{align}
This code construction lower bounds the DIF capacity in Eq.~\eqref{eq:double_capacity}:
\begin{align}
    C^\text{d}_\text{IDF}(\mathcal{P})
    &:= \inf_{\lambda_1,\lambda_2>0}\liminf_{n\rightarrow\infty}\frac{1}{n}\log\log N_{\text{IDF}}^\text{d}(K,\lambda_1,\lambda_2)\nonumber\\
    &\geq\lim_{n\to\infty}\frac{\log{\log{N}}}{n}\nonumber\\
    &=\lim_{n\to\infty}\frac{\sum_{k=0}^{K(n,\kappa)}H\left(\text{Pois}(p_k\hat{E}T_s+\lambda_0)\right)}{K(n,\kappa)+1}.\label{eq:approx}
\end{align}
Using the approximation of the entropy of Poisson distributed RV in \cite{elementsofIT}, and the assumption of memory $K(n,\kappa)=2^{\kappa\log n}$, we can approximate Eq.~{\eqref{eq:approx}} by:
\begin{equation*}
\begin{split}
&\lim_{n\to\infty}\frac{1}{K(n,\kappa)+1}\sum_{k=0}^{K(n,\kappa)}\left\{\frac{1}{2}\log\left[2\pi e\hat{E}T_s\left(p_k+\frac{\lambda_0}{\hat{E}T_s}\right)\right]\right.\nonumber\\
    &\quad\quad\quad\quad\left.-\frac{1}{12\cdot(p_k\hat{E}T_s+\lambda_0)}+\mathcal{O}\left(\frac{1}{(p_k\hat{E}T_s+\lambda_0)^2}\right)\right\}\nonumber\\
    &=\frac{1}{2}\log{\left(2\pi e \hat{E}T_s\right)}\lesssim  C^\text{d}_\text{IDF}(\mathcal{P}) ,\nonumber\\
\end{split}
\end{equation*}
completing the proof of Theorem \ref{thm:main2}.\hfill\qed

%============================= conclusion ====================================
\section{Conclusion}
\label{sec:conclusion}
In this paper, we studied DI over DTPCs with ISI, a setting that realistically captures memory effects in MC systems. By building on recent developments in identification theory, we significantly improved the previously known upper bound on the DI capacity under power constraints, tightening it from $\frac{3}{2}+\kappa$ to $\frac{1+\kappa}{2}$. Furthermore, we initiated the study of DIF for DTPCs with ISI, introducing a constructive lower bound based on a tailored coding scheme, and proving the double exponential nature of this protocol. These results contribute to a deeper theoretical understanding of event-driven communication in constrained MC environments. Future directions include refining the DIF capacity bounds, exploring identification under more complex feedback mechanisms, and analyzing practical coding strategies for finite block lengths in realistic MC applications.

%============================= acknowledgments ==============================
\section*{Acknowledgments}
%\yz{Please check the acknowledgments and add the acknowledgments for Pau.}
%\cd{Please use also our TUBS number for 6G-life}
%\yz{I have added our TUBS number. @Pau: Could you please add your acknowledgments?}
%\pau{I am unsure about my acknowledgments, I have asked Holger if there is anything I should add, but I think I am still fully supported by the IAS...}
H. Boche, C. Deppe, and Y. Zhao acknowledge financial support from the Federal Ministry of Education and Research of Germany (BMBF) through the program “Souverän. Digital. Vernetzt.” as part of the joint project 6G-life (project identification numbers: 16KISK002 and 16KISK263). Additionally, H. Boche received partial support from the BMBF under the national initiative on Post Shannon Communication (NewCom) through Grant 16KIS1003K. C. Deppe also received partial support under NewCom through Grant 16KIS1005. Furthermore, C. Deppe and Y. Zhao were supported by the DFG within the project DE1915/2-1.

\clearpage
\newpage
\bibliographystyle{IEEEtran}
\bibliography{definitions,references}
\clearpage
\newpage

\IEEEtriggeratref{4}
%%
%% which triggers a \newpage (i.e., new column) just before the given
%% reference number. Note that you need to adapt this if you modify
%% the paper.  The "triggered" command can be changed if desired:
%%
%\IEEEtriggercmd{\enlargethispage{-20cm}}
%%
%%%%%%
%\section*{References}
%\bibliographystyle{elsarticle-num}

%%%%%%
%% References:
%% We recommend the usage of BibTeX:
%%
%\bibliographystyle{IEEEtran}
%\bibliography{definitions,bibliofile}
%%
%% where we here have assume the existence of the files
%% definitions.bib and bibliofile.bib.
%% BibTeX documentation can be obtained at:
%% http://www.ctan.org/tex-archive/biblio/bibtex/contrib/doc/
%%%%%%

%% Or you use manual references (pay attention to consistency and the
%% formatting style!):

\end{document}